\documentclass[fleqn,usenatbib]{mnras}

\usepackage{newtxtext,newtxmath}

\usepackage[T1]{fontenc}
\usepackage{ae,aecompl}

\usepackage{graphicx}	
\usepackage{amsmath}	
\usepackage{amssymb}	

\newcommand{\nitrogen}{[N\,{\sc ii}]}

\newcommand{\nitrogena}{[N\,{\sc i}]}

\newcommand{\oxygeni}{[O\,{\sc i}]}
\newcommand{\oxygenii}{[O\,{\sc ii}]}

\newcommand{\sulfurt}{[S\,{\sc ii}]}

\def\vhel{\ifmmode{V_{{\rm HEL}}}\else{$V_{{\rm HEL}}$}\fi}
\def\vsys{\ifmmode{V_{\rm sys}}\else{$V_{\rm sys}$}\fi}
\def\kms{\ifmmode{~{\rm km\,s}^{-1}}\else{~km~s$^{-1}$}\fi}
\def\vlsr{\ifmmode{v_{\rm lsr}}\else{$v_{\rm lsr}$}\fi}

\usepackage[normalem]{ulem} 

\title[H$_2$ LISs in NGC~7009 and NGC~6543]{H$_2$ emission in the low-ionization structures of the Planetary Nebulae NGC~7009 and NGC~6543}

\author[Stavros Akras et al.]{
Stavros Akras$^{1,2,3}$\thanks{E-mail: stavrosakras@gmail.com},
Denise R. Gon\c{c}alves$^{1}$,
Gerardo Ramos-Larios$^{4}$,
Isabel Aleman$^{5}$
\\
$^{1}$Observat\'{o}rio do Valongo, Universidade Federal do Rio de Janeiro, Ladeira Pedro Antonio 43, 20080-090, Rio de Janeiro, Brazil\\
$^{2}$Instituto de Matem\'{a}tica, Estat\'{i}stica e F\'{i}sica, Universidade Federal do Rio Grande, Rio Grande 96203-900, Brazil\\
$^{3}$Observat\'orio Nacional/MCTIC, Rua Gen. Jos\'{e} Cristino, 77, 20921-400, Rio de Janeiro, Brazil\\
$^{4}$Instituto de Astronom\'{i}a y Meteorolog\'{i}a, CUCEI, Av. Vallarta No. 2602, Col. Arcos Vallarta, CP 44130, Guadalajara, Jalisco, Mexico\\
$^{5}$Instituto de F\'isica e Qu\'imica, Universidade Federal de Itajub\'a, Av. BPS 1303 Pinheirinho, 37500-903 Itajub\'{a}, MG, Brazil
}

\date{Accepted XXX. Received YYY; in original form ZZZ}

\pubyear{2015}

\begin{document}
\label{firstpage}
\pagerange{\pageref{firstpage}--\pageref{lastpage}}
\maketitle

\begin{abstract}

Despite the many studies in the last decades, the low-ionization structures (LISs) of planetary nebulae (PNe) still hold several mysteries. Recent imaging surveys have demonstrated that LISs are composed of molecular gas. Here, we report H$_2$ emission in the LISs of NGC~7009 and NGC~6543 by means of very deep narrow-band H$_2$ images taken with NIRI@Gemini. The surface brightness of the H$_2$ 1-0 S(1) line is estimated to be (0.46--2.9)$\times$10$^{-4}$ erg~s$^{-1}$~cm$^{-2}$~sr$^{-1}$ in NGC 7009 and (0.29--0.48)$\times$10$^{-4}$ erg~s$^{-1}$~cm$^{-2}$~sr$^{-1}$ in NGC 6543, with signal-to-noise ratios of 10-42 and 3-4, respectively. These findings provide further confirmation of hidden H$_2$ gas in LISs. The emission is discussed in terms of the recent proposed diagnostic diagram $R$(H$_2$)=H$_2$~1-0~S(1)/H$_2$~2-1~S(1) versus $R$(Br$\gamma$)=H$_2$~1-0~S(1)/Br$\gamma$, which was suggested to trace the mechanism responsible for the H$_2$ excitation. Comparing our observations to shock and ultraviolet (UV) molecular excitation models, as well as a number of observations compiled from the literature showed that we cannot conclude for either UV or shocks as the mechanism behind the molecular emission.

\end{abstract}




\section{Introduction}

Thirty two years have passed since the report of the `low-ionization inclusions' in planetary nebulae (PNe) by \citet{Balick1987}. After that pioneering work, several studies have been carried out, using high-quality imagery and spectroscopy, with the aim to unveil the true nature of these microstructures, explore the physical properties that make them differ from the surrounding nebular medium, in terms of the emission line fluxes, and give insights into their formation mechanism \citep[e.g.][]{Balick1993,Balick1994,Balick1998,Hajian1997}.

Since then, several PNe have been found to host these microstructures \citep[e.g.][]{Corrdi1996} and different labels have been given based on their  properties: fast low-ionization emission regions \citep[FLIERs, ][]{Balick1993}, slow moving low ionization emitting regions \citep[SLOWERs, ][]{Perinotto2000} or bipolar, rotating, episodic jets \citep[BRETs, ][]{Lopez1993,Lopez1995}. We will hereafter refer to all of them as low-ionization structures or LISs \citep[see ][]{Goncalves2001}.

\begin{table*}
\caption{Observations log} 
\centering
\begin{tabular}{lcccccc}
\hline
\hline
\multicolumn{1}{c}{Filter} & \multicolumn{1}{c}{$\rm{\lambda_c}$} &
\multicolumn{1}{c}{$\rm{\Delta\lambda}$} & \multicolumn{1}{c}{Time$^{a}$} & \multicolumn{1}{c}{ Number of frames} & \multicolumn{1}{c}{Time$^{a}$} & \multicolumn{1}{c}{ Number of frames} \\
      &  ($\micron$) & ($\micron$) & (s) & & (s) & \\
\hline
\multicolumn{1}{c}{} & \multicolumn{1}{c}{} & \multicolumn{1}{c}{} & \multicolumn{2}{c}{NGC~7009} & \multicolumn{2}{c}{NGC~6543}\\
\hline
K-cont-1    & 2.0975 & 0.0275 &  100  & 7  &  90 & 7 \\
H$_2$ 1-0 S(1) & 2.1239 & 0.0261 &  100  & 6 &  90 & 7 \\
Brackett~$\rm{\gamma}$  & 2.1686 & 0.0295 &  45  & 5 &  45  & 5 \\ 
H$_2$ 2-1 S(1) & 2.2465 & 0.0301 &  173 & 17 &  163 & 16 \\
K-cont-2    & 2.2718 & 0.0352 &  173 & 16 &  163 & 14 \\
\hline
\multicolumn{7}{l}{$^{a}$ Integration time of each individual frame.}
\end{tabular}
\end{table*}

\citet{Goncalves2001} reviewed the kinematic and morphological characteristics of LISs and discussed possible links with various formation models. The authors came to the conclusion that there is no direct connection between LISs and the morphology of PNe, since they appear in all types (round, elliptical, bipolar, or point-symmetric). Various formation models of knots and/or jets are able to explain some of their observable characteristics but not all of them. So far, there is not a general model that can provide an adequate explanation for all the microstructures.

Shock interactions have been proposed as a possible mechanism to explain the enhancement in low-ionization lines like \oxygeni, \nitrogena, \sulfurt, \oxygenii, and \nitrogen\ \citep[e.g.][]{Hartigan1994,dopita1997}. Running a series of numerical simulations for a high-density knot moving outwards in a less dense medium, \citet{Raga2008} demonstrated that the spectral characteristics of shock-excited or photoionized regions can both be reproduced by changing the local photoionization rate. Low photoionization rate models generate spectroscopic characteristics similar to shock-excited regions. A crucial parameter to distinguish the mechanisms is the distance between the LISs and the source of ionizing photons, which determines the photoionization rate at the position of the structures \citep[][]{Aleman_knots_2011,Akrasden2016}.

Based on Raga's simulations and spectroscopic data from a sample of PNe with LISs, \citet{Goncalves2009} argued that the spectra of high-velocity knots can be explained by shocks. A few years later, \cite{Akrasden2016} shown that the enhancement of low-ionization lines in knots relative to the surrounding medium is the result of a combination of the ultraviolet (UV) radiation from the central star and shock interactions. However, it is not easy to distinguish the contribution from each mechanisms. On the other hand, there are studies which attribute the enhancement of low-ionization lines in some LISs only to the UV stellar radiation of the central star \citep[e.g.][]{Hajian1997,Denise2004,Ali2017}. Despite all this effort so far, LISs are still poorly understood. 

One of the most intriguing characteristic of LISs is the electronic density (determined from the common diagnostic line ratios), which is systematically lower than or at most equal to the electronic density of the surrounding nebular gas \citep[e.g.][]{Balick1993,Hajian1997,Goncalves2003,Goncalves2009,Monteiro2013,Akrasden2016,Ali2017}. This finding contradicts the formation models of knots in which they are considered more dense than the surrounding nebular gas \citep[e.g.][]{Steffen2001,Raga2008}. \citet{Goncalves2009} proposed the scenario that LISs are also made of molecular gas and dust, similarly to the cometary knots of Helix~\citep[e.g.][]{Huggins2002,Meixner2005,Matsuura2007,Matsuura2009}. 

\cite{Matsuura2007,Matsuura2008} shown that the intensities of the ro-vibrational H$_2$ lines from the cometary knots of Helix can be explained either by a low-velocity shock of 27~km~s$^{-1}$ or a strong UV stellar radiation field. Interestingly, the latter requires a more luminous central star than the observations indicate i.e. higher local ionization parameter. More detailed simulations by \citet{Aleman_knots_2011} have, however, shown that shocks are not needed to explain the observed H$_2$ surface brightness.

Besides, the cometary of Helix, H$_2$ emission has also been detected in the cometary knots of the Ring nebula \citep[][]{Speck2003} and the Dumbbell nebula \citep[][]{baldridge2017}.

\begin{table*}
\begin{center}
\caption{Emission-line fluxes for NGC~7009 and NGC~6543, in units of 10$^{-15}$ erg~s$^{-1}$~cm$^{-2}$. $R$(H$_2$) = H$_2$ 1-0~S(1)/1-2~S(1) and $R$(Br$\gamma$) = H$_2$~1-0~S(1)/Br$\gamma$. Numbers in parenthesis correspond to the S/Ns. Lower rows give the surface brightness in units of  10$^{-4}$~erg~s$^{-1}$~cm$^{-2}$~sr$^{-1}$.} 
\begin{tabular}{lcccccccc}
\hline
\hline
\noalign{\smallskip}
Name & \multicolumn{1}{c}{R.A.} & \multicolumn{1}{c}{Dec.} &
\multicolumn{1}{c}{H$_2$~1-0~S(1)} & \multicolumn{1}{c}{H$_2$~2-1~S(1)} & \multicolumn{1}{c}{Br$\rm{\gamma}$}
& $R$(H$_2$) & $R$(Br$\rm{\gamma)}$ & \multicolumn{1}{c}{Box (arcsec$^2$)} \\
\hline
\multicolumn{9}{l}{NGC 7009:}\\
\\
LIS-E1  &  21:04:12.486 & -11:21:41.903 & 3.71 (42) & 0.38 (19) & 2.61 (17) &  9.76  & 1.42 & 0.583~$\times$~0.931\\
    &               &               & 2.91       & 0.30    & 2.05     &        &       & \\ 
LIS-E2  &  21:04:12.743 & -11:21:36.836 & 1.06 (10) & 0.13 (5) & 2.05 (9) &  8.15  & 0.52 & 0.932~$\times$~1.049\\
    &               &               & 0.46       & 0.06    & 0.89     &        &       & \\ 
LIS-W1  &  21:04:09.029 & -11:21:52.942 & 11.4 (14) & 3.26 (8) & 28.2 (11) &  3.50  & 0.40 & 3.965~$\times$~2.564\\
 &               &               & 0.48       & 0.14    & 1.18     &        &       & \\ 
LIS-W1  &  21:04:09.029 & -11:21:53.175 & 8.13 (15) & 2.39 (10) & 16.7 (12) &  3.40  & 0.49 & 2.682~$\times$~2.214\\
 &               &               & 0.58       & 0.17    & 1.20     &        &       & \\ 
LIS-W1  &  21:04:09.009 & -11:21:52.933 & 7.36 (15) & 1.98 (7) & 13.6 (11) &  3.72  & 0.54 & 3.382~$\times$~1.515\\
 &               &               & 0.61       & 0.16    & 1.13     &        &       & \\ 
\hline
\multicolumn{9}{l}{NGC 6543:}\\
\\
LIS-SW  & 17:58:32.103 & +66:37:47.725  & 2.99 (3) & <0.9$^a$  & 44.8 (7) & >3.32$^b$  & 0.07 & 2.020~$\times$~1.858\\
 &               &               & 0.34      & <0.10$^a$    & 5.08     &        &       & \\ 
LIS-SW  & 17:58:32.053  & +66:37:47.404  & 6.03 (3) & <2.08$^a$  & 86.3 (7) & >2.89$^b$  & 0.07 & 3.150~$\times$~2.760\\
 &               &               & 0.30      & <0.10$^a$    & 4.22     &        &       & \\ 
LIS-SW  & 17:58:32.155  & +66:37:46.968  & 1.26 (3) & <0.44$^a$ & 8.56 (6) &  >2.86$^b$  & 0.15 & 0.747~$\times$~2.468\\
 &               &               & 0.29      & <0.10$^a$    & 1.98     &        &       & \\ 
 LIS-NE  & 17:58:33.800  & +66:38:11.717  & 1.97 (4) & <0.42$^a$ & 35.1 (7) & >4.69$^b$  & 0.06 & 1.723~$\times$~1.002\\
 &               &               & 0.49      & <0.10$^a$    & 8.65     &        &       & \\ 
\hline
\multicolumn{9}{l}{$^a$These numbers correspond to the 3$\sigma$ upper limits. $^b$These numbers correspond to the 3$\sigma$ lower limits.}
\end{tabular}
\end{center}
\end{table*}

None the less, H$_2$ emission had not been detected in LISs despite various surveys \citep[e.g.][]{Latter1995,Kastner1996,Hora1999,Guerrero2000}. The main reason for that was the limited spatial resolution and sensitivity of these observations. More sensitive and deeper observations over the last five years have already provided strong evidence as well as the first direct confirmations of H$_2$ emission associated with LISs in PNe. First, \cite{Fang2015} detected H$_2$ emission from the northwestern knot of Hu~1-2. Couple of years later, Akras, Gon\c calves and Ramos-Larios (2017) succeed to detect the H$_2$ emission from the LISs in two PNe, K~4-47 and NGC~7662, using very deep narrow-band imagery from the 8~m Gemini North telescope. \cite{Fang2018} reported the detection of H$_2$ emission from two distant pairs of knots in Hb~12, from a number of randomly distributed knots in the haloes of NGC~6543 and NGC~7009 as well as from the pair of LISs in the ionized region of NGC~7009. Regarding Hb~12, it should be mentioned that despite the H$_2$ from its distant clumps, strong H$_2$ 1-0 S(1) emission has also been detected in the centre of the nebula \citep{Dinerstein1988,Ramsay1993,Luhman1996}, and its origin is pure UV-fluorescence. It should also be noted that H$_2$ emission detected in the equatorial regions of some bipolar PNe has been unveiled to be fragmented into clumps and filaments \citep{Marquez2013,Manchado2015}. 

In this paper, we present new deep H$_2$ narrow-band NIRI@Gemini images for two of the most well-studied PNe, NGC~7009 and NGC~6543. The paper is organized as follows: observations are described in Section~2. In Sections~3 and 4, we present the results of H$_2$ detection in NGC~7009 and NGC~6543. The mechanisms responsible for the excitation of molecular hydrogen in these structures is discussed in Section 5, and we finish with our conclusions in Section~6.

\begin{figure*}
\includegraphics[scale=0.57]{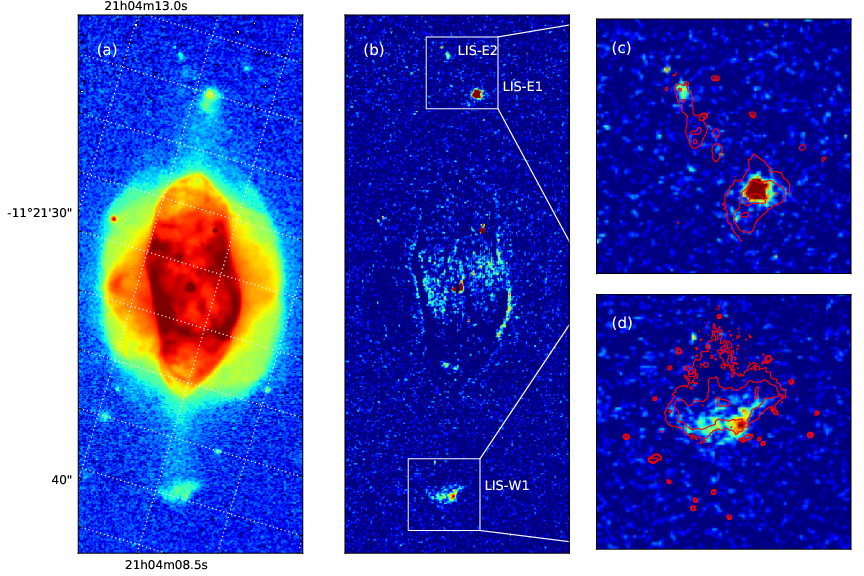}
\caption[]{Gemini NIRI images of NGC~7009. LISs are labelled as W1, E1 and E2. H$_2$~1-0~S(1) line image (panel a), H$_2$~1-0~S(1) continuum-subtracted image (panel b), and H$_2$~1-0~S(1) continuum-subtracted images of W1 and E1/E2 LISs overlaid by {\it HST} \nitrogen~$\lambda$6584 emission contours (panels c and d).}
\label{NGC7009H21}
\end{figure*}

\section{Observations} 
\label{sec:obs} 

The observations of NGC 7009 and NGC 6543 were acquired in service mode on 2017 May 10 and June 24 (Program ID: GN-2017A-Q-58, PI: S. Akras) using the Near InfraRed Imager and Spectrometer on the Gemini-North Telescope at Mauna Kea in Hawaii.

The narrow-band filters G0216, G0218 and G0220 were used to isolate the H$_2$~1-0~S(1), H$_2$~2-1~S(1) and Br$\gamma$ lines centred at 2.1239, 2.2465 and 2.1686~$\mu$m, respectively, as well as the filters G0217 and G2019 centred at 2.0975 and 2.2718~$\mu$m  to obtain the continuum emission. Due to the targets' sizes, the f/6 configuration was adopted providing a field of view of 120~arcsec$^2$ and pixel size of 0.117~arcsec. Several individual frames, with different exposure times per filter, were obtained in order to increase the signal-to-noise (S/N) ratio. The observing log is summarized in Table 1.

Darks and GCAL flat frames were also obtained for the correction of the thermal emission, dark current and hot pixels. In order to reduce the total time of the observations, the major axis of the targets was oriented in the up-down direction on the detector, which made possible to carry out the dithering across the left-right direction (minor-axis of the nebula) and avoid the additional observations for the sky background. Standard stars (TYC~4413-304-1 and GSPC~S813-D) were also observed to flux calibrate the data.

All first frames from each sequence were removed from the reduction/analysis as recommended by Gemini. Before start the reduction process, the {\sc python} routines CLEARIR.py and NIRLIN.py\footnote{CLEARIR.py and NIRLIN.py {\sc python} routines were developed and are distributed by Gemini for a preparatory reduction of the data obtained with NIRI and GNIRS, https://www.gemini.edu/sciops/instruments/niri/} were applied to all the frames for the correction of the vertical stripping and the non-linearity of the detector. The reduction of the data was then performed with the GEMINI \textsc{iraf}\footnote{ {\sc iraf} is distributed by the National Optical Astronomy Observatories, which are operated by the Association of Universities for Research in Astronomy, Inc., under cooperative agreement with the National Science Foundation} package for NIRI. The routines {\it Nprepare, Nisky, Niflat, Nireduce} and finally {\it Imcoadd} were used for each imaging set accordingly.

The estimated surface brightness sensitivity of the H$_2$~1-0~S(1), H$_2$~2-1~S(1) and Br$\gamma$ images are 2.8$\times$10$^{-16}$, 1.3$\times$10$^{-16}$, and 1.9$\times$10$^{-15}$  erg~s$^{-1}$~cm$^{-2}$~arcsec$^{-2}$, respectively for NGC~7009. For NGC~6543, the corresponding sensitivities are 3.1$\times$10$^{-16}$, 2.4$\times$10$^{-16}$ and ~2.3$\times$10$^{-15}$ erg~s$^{-1}$~cm$^{-2}$~arcsec$^{-2}$, respectively.


\section{H$_2$ in NGC~7009}

In spite of NGC~7009 being one of the most well-studied PN, only a year ago H$_2$ emission was reported at the tips of the two jet or jet-like LISs in the ionized region of this nebula \citep{Fang2018}. However, these detections were not supported by a continuum-subtracted image or any H$_2$~1-0~S(1) flux measurement, but only by the spatial offset between the H$_2$ and Br$\gamma$ emission-lines. Therefore, more observations were needed to confirm the results.

Before the undoubted detection of H$_2$ gas in the LISs of NGC~7009, \citet{Phillips2010} had carried out a detailed analysis of this PN using Infrared Array Camera (IRAC) images (i.e. [3.6], [4.5], [5.8] and [8] $\mu$m) and the available spectra in the ISO and {\it Spitzer} archives. From the three-colours RGB IRAC image of NGC~7009 \citep[see fig.~1 in][]{Phillips2010}, both LISs (outflow+knots) are displayed with green colour, which corresponds to the emission from the [4.5] band. Sources with enhanced emission at the [4.5] band are usually found close to young stellar objects and attributed to outflows \citep[]{DeBuizer2010}, being known as \lq\lq Green fuzzies\rq\rq\ \citep[]{Chambers2009}. By analysing spectroscopic data of \lq\lq Green fuzzy\rq\rq\ sources, \citet{DeBuizer2010} argued that H$_2$ emission is likely the primary emission in the [4.5] band. The same correlation has also been found in the Helix nebula, for which the cometary knots are unambiguously bluer in the [4.5]-[8.0] colour index compared to the nebular gas due to the stronger emission from the [4.5] band dominated by H$_2$ lines \citep[][]{Hora1999}. Molecular hydrogen lines from the ground rotational state such as 0-0~S(7), 0-0~S(4), and 0-0~S(5) may also contribute to the [5.8] and [8.0] IRAC bands \citep{Hora2004}. The strong IRAC [4.5] emission found in the outflow and knots of NGC~7009 was an indirect indication of possible H$_2$ emission in these substructures, but a strong contribution from the Br$\alpha$ line at 4.052~$\mu$m could not be ruled out \citep{Phillips2010}.

\begin{figure*}
\includegraphics[scale=0.57]{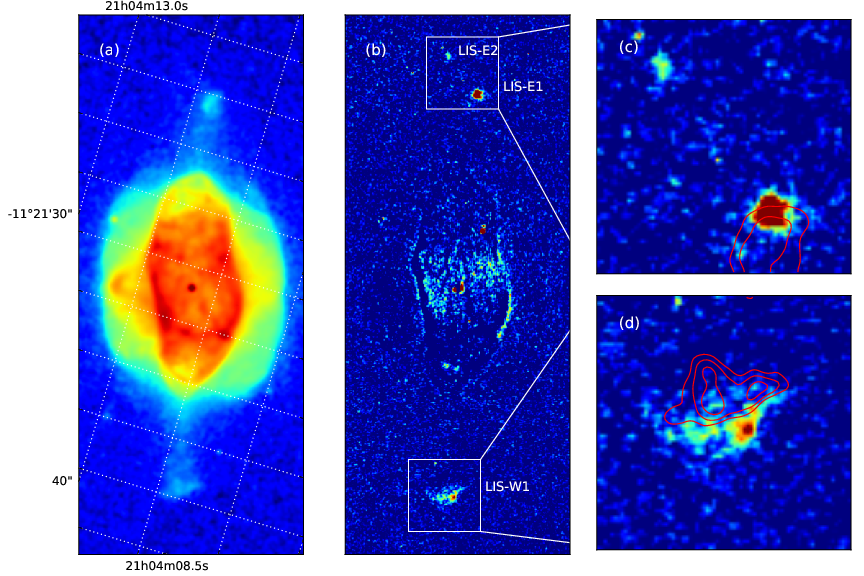}
\caption[]{The same as Figure~\ref{NGC7009H21}. Br$\gamma$ line image (panel a), H$_2$~1-0~S(1) continuum-subtracted image (panel b), and H$_2$~1-0~S(1) continuum-subtracted images of W1 and E1/E2 LISs overlaid by Br$\gamma$ emission contours (panels c and d).}
\label{NGC7009Brg}
\end{figure*}

\begin{figure*}
\includegraphics[scale=0.57]{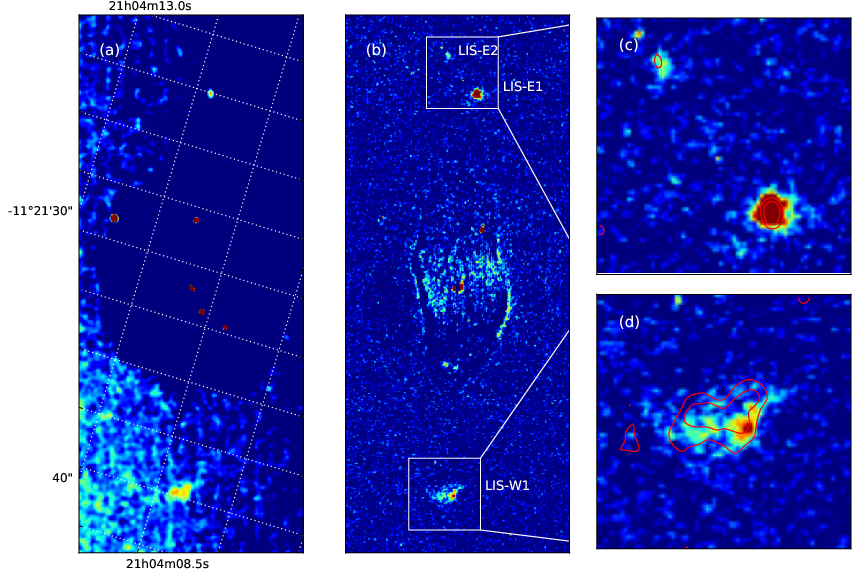}
\caption[]{The same as Figure~\ref{NGC7009H21}. H$_2$~2-1~S(1) continuum-subtracted image (panel a), H$_2$~1-0~S(1) continuum-subtracted image (panel b), and H$_2$~1-0~S(1) continuum-subtracted images of W1 and E1/E2 LISs  overlaid by H$_2$~2-1~S(1) emission contours (panels c and d).}
\label{NGC7009H22}
\end{figure*}

Our more sensitive and high spatial resolution narrow-band images of NGC~7009 have detected the ro-vibrational H$_2$ 1-0~S(1) and 2-1~S(1) emission-lines from three substructures, confirming the results from \citet[]{Fang2018}. They are designated E1, E2, and W1 in Figures~\ref{NGC7009H21}-\ref{NGC7009H22}. The main body of NGC~7009 is characterized by an ellipsoidal nebula, known as Saturn nebula, with a pair of highly collimated outflows and knots. Both knots are bright in low-ionization lines such as \nitrogen, \oxygenii, \sulfurt, and \oxygeni\ \cite[][]{Goncalves2003,Goncalves2006,Walsh2018}.  The E1 and W1 LISs correspond to the K1 and K4 knots in \citet[][]{Goncalves2003,Goncalves2006}. There is also a marginal detection of H$_2$ emission from the main nebula but only spectroscopic follow-up observations can confirm whether it is real or remnant from the continuum subtraction process. It is interesting, though, that this emission follows the morphology of the outer edge of the ellipsoidal nebula.

The emission line fluxes of the three LISs are listed in Table~2. The H$_2$~1-0~S(1) line flux varies from 1.06 to 11.4$\times$10$^{-15}$ erg~s$^{-1}$~cm$^{-2}$ with an S/N higher than 10, while the H$_2$~2-1~S(1) line flux varies between 0.129 and 3.26$\times$10$^{-15}$ erg~s$^{-1}$~cm$^{-2}$ with an S/N higher than 5. The $R$(H$_2$)=H$_2$~1-0~S(1)/H$_2$~1-2~S(1) line ratio is found to be significantly different between the eastern and western LISs. In particular, the E1 and E2 LISs have $R$(H$_2$) equal to 9.81 and 8.22, respectively, which are approximately three times higher than the value of W1 LIS. Because of the irregular morphology of W1 LIS, we present three estimates of the fluxes obtained from different aperture sizes, and all of them result in very similar $R$(H$_2$) ratio, from 3.40 to 3.72.

Br$\gamma$ emission is also detected in all the three LISs with fluxes between 2.05 and 28.2 $\times$10$^{-15}$ erg~s$^{-1}$~cm$^{-2}$. The $R$(Br$\gamma$)=H$_2$~1-0~S(1)/Br$\gamma$ line ratio does not show significant differences among the LISs (see Table~1), and it is comparable with the values derived for the LISs in NGC~7662 \citep[][]{Akras2017}.
 
A comparison between the H$_2$~1-0~S(1) and H$_2$~2-1~S(1) lines with the \nitrogen\ {\it Hubble Space Telescope (HST)} image at 6584~\AA\ shows a very good match (Figures~\ref{NGC7009H21} and \ref{NGC7009H22}), with the former being engulfed by the \nitrogen\ emission. Figure~\ref{NGC7009Brg} displays the distribution of the Br$\gamma$ line and its comparison with the bright H$_2$~1-0~S(1) line. The spatial displacement between the Br$\gamma$ and H$_2$ emissions reported by \citet{Fang2018} is also observed in our data.

\citet[][]{Goncalves2003,Goncalves2006} identified two more LISs, labelled K2 and K3, which lie closer to the central star, but no H$_2$ emission is detected in these two LISs. The detection of H$_2$ emission in four more knots distributed in the halo of NGC~7009, at distances higher than 1~arcmin from the central star, has also been reported \citep[]{Fang2018}.

\section{H$_2$ in NGC~6543}

NGC~6543, the second PN observed with NIRI at the Gemini-North   (Program ID:GN-2017A-Q-58), is characterized by a bipolar structure with a pair of condensations and jets bright in low ionization lines \citep{Balick1994,Balick2004,balickHajian2004}. 

Similar to NGC~7009, halo features in NGC~6543 also appear with green colour in the three-colours RGB IRAC images \citep[fig.~1 and 9 in][]{Hora2004,Fang2018}. As we have already pointed out, the strong emission from the [4.5] band is attributed to molecular lines that fall within this band \citep[][]{Hora1999,DeBuizer2010}. Only last year, Fang and coworkers reported the detection of H$_2$ emission from the knotty and filamentary halo that surrounds NGC~6543 \citep{Fang2018}.

\begin{figure*}
\includegraphics[scale=0.57]{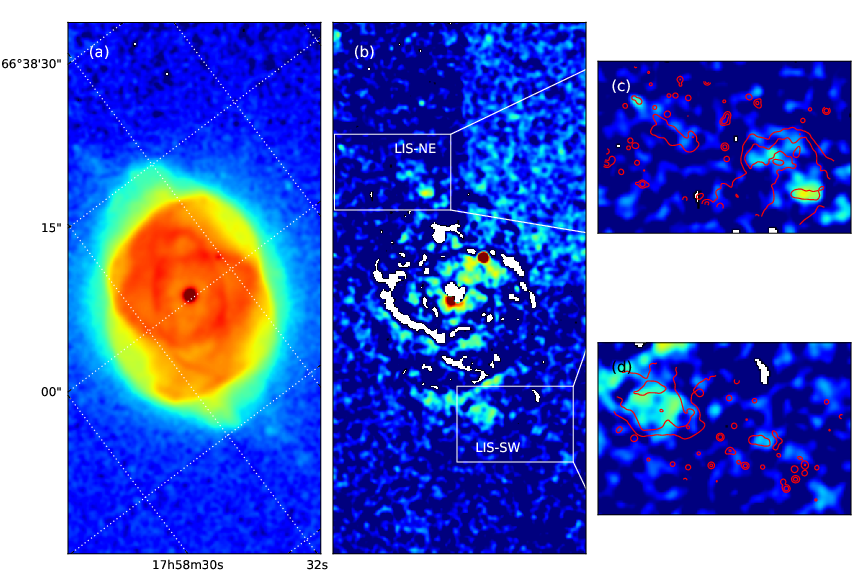}
\caption[]{Gemini NIRI imaging of NGC~6543. The pairs of LISs are labelled as SW and NE. H$_2$~1-0~S(1) line image (panels a), H$_2$~1-0~S(1) continuum-subtracted image smoothed using a Gaussian function with a radius of 2~$\sigma$ (panel b), and H$_2$~1-0~S(1) continuum-subtracted images of SW and NE LISs overlaid by {\it HST} \nitrogen~$\lambda$6584 emission contours (panels c and d).}
\label{NGC6543H21}
\end{figure*}

In contrast to Fang et al.'s (2018) work, the focus of this work is on the bright ionized bipolar nebula of NGC~6543, also known as the Cat's Eye Nebula, and not on its recombination halo. The narrow-band H$_2$~1-0~S(1) NIRI image of NGC~6543 is presented in Figure~\ref{NGC6543H21}. The image is smoothed using a Gaussian function with a radius of 2~$\sigma$ so as we can better illustrate the detection of the emission in the SW and NE LISs. These LISs correspond to the F-F$^{\prime}$ condensations between the caps and the jets in \citet{Miranda1992}. The H$_2$~1-0~S(1) line fluxes are estimated between 1.26 and 2.99$\times$10$^{-15}$ erg~s$^{-1}$~cm$^{-2}$ with S/N ratio of 3-4 (see also Table~2). Three flux measurements derived from different aperture sizes are given for the SW LIS. The detection of H$_2$ emission from both LISs of NGC~6543 is reported for the first time. A marginal detection of H$_2$ emission at the edges of the bipolar lobes, labelled as \lq\lq caps\rq\rq\ in \cite{Balick1994} and D-D$^{\prime}$ in \cite{Miranda1992} can also be seen in Figure~\ref{NGC6543H21}, but deep follow-up spectroscopic observations are necessary to confirm this emission. The H$_2$~2-1~S(1) line at 2.24~$\mu$m is not detected in NGC~6543, but we obtained upper limits for its emission, which yields lower limits for the $R$(H$_2$) ratio between 2.9 and 4.7 (Table~2).

\begin{figure*}
\includegraphics[scale=0.57]{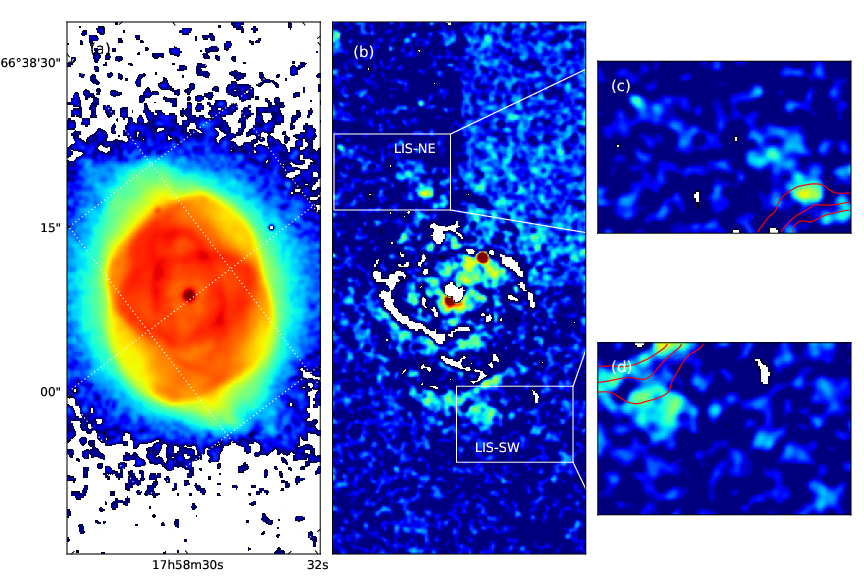}
\caption[]{The same as Figure~\ref{NGC6543H21}. Br$\gamma$ line image (panel a), H$_2$~1-0~S(1) continuum-subtracted image smoothed using a Gaussian function with a radius of 2~$\sigma$ (panel b), and H$_2$~1-0~S(1) continuum-subtracted images of SW and NE LISs overlaid by Br$\gamma$ emission contours (panels c and d).}
\label{NGC6543Br}
\end{figure*}

Both LISs in NGC~6543 are detected in the Br$\gamma$ line (Figure~\ref{NGC6543Br}) with fluxes between 8.56 and 86.3$\times$10$^{-15}$ erg~s$^{-1}$~cm$^{-2}$ and S/N ratio of 6-7 (see Table~2). Their $R$(Br$\gamma$) ratios have been estimated between 0.06 and 0.15, close to the values calculated for some LISs in NGC~7662 \citep{Akras2017}. From Figure~\ref{NGC6543Br}, it can also be seen that a diffuse emission from the nebula itself surrounds the LISs and it may contribute to the Br$\gamma$ fluxes. This additional emission may be responsible for the very low $R$(Br$\gamma$) ratios found in the LISs.

A comparison of our H$_2$~1-0~S(1) image with the {\it HST} \nitrogen~$\lambda$6584 image show a strong correlation (Figure~\ref{NGC6543H21}). Similarly to NGC~7009, H$_2$ emission is engulfed by the more extended \nitrogen\ emission. Comparison of the spatial distribution in H$_2$ and Br$\gamma$ emission-lines illustrates an offset between the two emissions as in NGC~7009.

\section{Molecular hydrogen excitation}

The mechanisms to populate the H$_2$ (v,J)~=~(1,3) and (2,3) rovibrational levels, whose decay yield the emission in the 1-0~S(1) and 2-1~S(1) line, respectively, may be dominated by UV pumping or collisions with gas particles, depending on the local physical conditions \citep[see][ and references therein]{Aleman2011}. For these levels, direct radiative excitation and de-excitation are forbidden via dipole transitions, but allowed by electric quadrupole. The latter has low probability of occurrence, but is the decay mechanism that produces the lines we observe. Formation pumping is also possible, but it is not an important mechanism for those levels in PNe conditions.

\begin{figure*}
\includegraphics[scale=0.565]{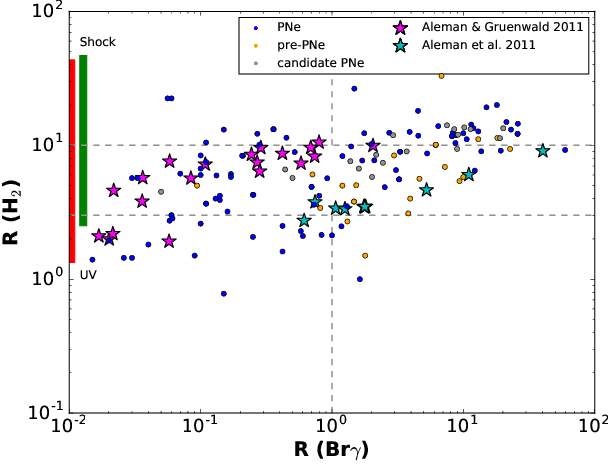}
\includegraphics[scale=0.565]{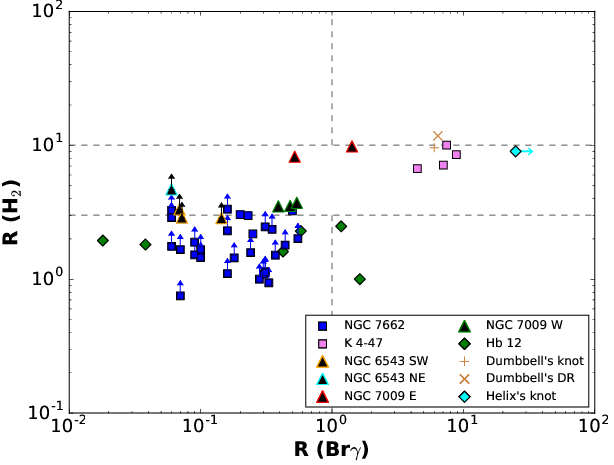}
\caption[]{Molecular hydrogen line ratio diagnostic diagram. Top panel: $R$(H$_2$) vs. $R$(Br$\gamma$) diagram for pre-PNe (orange circles), PNe (blue circles), and candidate PNe (gray circles) from the literature, theoretical predictions from UV models (ping/cyan stars and red vertical strip) and shock models (green vertical strip). Bottom panel: $R$(H$_2$) vs. $R$(Br$\gamma$) diagram for LISs in NGC~7009, NGC~6543, NGC~7662, and K~4-47 PNe, as well as Dumbbell's knot and nebular diffuse region, and Hb~12's central region (see the text for details). The horizontal dashed lines correspond the typical values for non-thermal ($\sim$3) and thermal emission ($\sim$10). The vertical dashed line indicates the border ($R$(Br$\gamma$)=1) between sources brighter in Br$\gamma$ (left-hand part) and brighter in H$_2$ 1-0.}
\label{plot}
\end{figure*}

Early works on H$_2$ excitation in PNe \citep[e.g. ][]{Beckwith1980} suggested the $R$(H$_2$) ratio as a diagnostic for the molecular excitation mechanism. $R$(H$_2$)$\sim$2 would indicate that H$_2$ is excited by UV radiation, while $R$(H$_2$)$>$4 would pointed out to H$_2$ gas excited by collisions with gas particles in shock regions. This inference was based on the sets of models available for UV excitation \citep[e.g.][]{black1976,black1978,shull1978}, and for shocks \citep[][]{Kwan1977,shullhollenbach1978}. \citet{Sternberg1989} showed, however, that the real picture was not that simple. In certain conditions, such as high-density gas irradiated by UV, collisional excitation can dominate the level excitation and give $R$(H$_2$) ratios that resemble those of a thermalized shock-heated gas.

\cite{Marquez2015} proposed as diagnostic for UV- and shock-excited H$_2$ gas in PNe a plot between the $R$(H$_2$) and $R$(Br$\gamma$) line ratios. Yet, these two line ratios are not enough to argue for the excitation mechanism of molecular hydrogen. For a better diagnostic analysis, it is necessary to study the H$_2$ level population of more levels, having in mind that all the mechanisms may compete in importance and that the dominant mechanism may vary for different levels and in different physical conditions \citep{Aleman2011}. Excitation diagrams are helpful tools to determine the excitation mechanism of H$_2$ \citep[e.g., ][]{Hora1999,Lumsden2001}. However, for narrow-band imaging studies like ours, it is not feasible to measure several lines and construct such diagrams. As we can derive the $R$(H$_2$) and $R$(Br$\gamma$) ratios from our observations, we construct the diagnostic diagram as suggested by \cite{Marquez2015} to make a preliminary assessment of which excitation mechanism dominates the H$_2$ level population in LISs. Follow-up spectroscopy studies will be necessary in order to construct more conclusive diagnostic diagrams.

Our $R$(H$_2$) vs. $R$(Br$\gamma$) diagnostic diagram is shown in Figure~\ref{plot}. In the top panel, we show pre-PNe (orange circles), PNe (blue circles), and candidates PNe (gray circles) gathered from the literature \citep{Geballe1991,aspin1993,Vicini1999,Hora1999,Lumsden2001,Garcia2002,Davis2003,Kelly2005, Likkel2006,Ramos2008,Gledhill2015,Marquez2015, Jones2018}. Model predictions of the $R$(H$_2$) ratio obtained from UV irradiated cloud models \citep[photodissociation regions; ][]{Sternberg1989,Burton1990} and shock models \citep[][]{Smith1995,Novikov2018} are indicated as red and green vertical stripes, respectively. 

From a quick observation of the diagnostic diagram, it can be seen that there is a trend between $R$(H$_2$) and $R$(Br$\gamma$) line ratios, both increasing from the lower right to the upper left corner. This behaviour was attributed to the transition from UV-dominated regions (left-hand side) to shock-dominated regions (right-hand side) \citep{Marquez2015}. But, according to the shocks and UV-irradiated cloud models, both mechanisms can provide a reasonable explanation for the observable range of $R$(H$_2$) values ($\sim$2-45).

Note that all but three pre-PNe exhibit $R$(Br$\gamma$)$>$1, while PNe cover the whole range of $R$(Br$\gamma$) line ratio from 0.01 up to 100. On the other hand, both PN and pre-PNe shows the same range of $R$(H$_2$) from $\sim$2 up to $\sim$20 (Figure~\ref{plot}). The high $R$(H$_2$) found in pre-PNe (or some PNe) indicates that the emission is thermalized, while for those with low $R$(H$_2$) the H$_2$ emission is non-thermal. These findings are consistent with the conclusions of \cite{Davis2003} that shocks may play an important role in the molecular gas excitation of pre-PNe and PNe \citep[see also][]{Akras2017,Akrasden2016}. However, it is not possible to trace  and disentangle the contribution of each excitation mechanism (UV fluorescence or shocks), with this diagnostic diagram. 

Two more sets of models, one for PNe \citep{Aleman2011} and one for the Helix nebula cometary knots \citep{Aleman_knots_2011} are also presented. Both sets correspond to self-consistent photoionization plus photodissociation region models, with the former corresponding to uniform density PNe, while the latter models were specifically developed to simulate one cometary knot (dense clump) subject to different conditions inside the Helix nebula. It is worth mentioning that for the three cometary knot models with $R$(Br$\gamma$)$\geq$2, the distance from the central source varies and the model with the largest distance predicts highest ratio. The importance of the distance between the dense knot and the central ionization source becomes clear from these models, as discussed in \citet{Aleman2011}, \citet{Akrasden2016} and \citet{Akras2017}.

The $R$(H$_2$) and $R$(Br$\gamma$) ratios of the LISs in NGC~7009 and NGC~6543 obtained in this work as well as those from  NGC~7662 and K~4-47 published in \citet{Akras2017} are shown in the bottom panel of Figure~\ref{plot}. A few other measurements obtained for PNe knots are also added in the plot. Studies of the H$_2$ emission from the PN Hb~12 (green diamonds; \citealt{Dinerstein1988}, \citealt{Ramsay1993}, \citealt{Luhman1996}) are also included in order to show the $R$(H$_2$) and $R$(Br$\gamma$) ratios for a molecular gas excited by UV stellar radiation. The $R$(H$_2$) ratio of the K1 cometary knot in Helix determined by \citet{Matsuura2007} is indicated with a cyan diamond. Br$\gamma$  emission is not detected in this knot and a lower limit is provided for the $R$(Br$\gamma$), assuming as upper limit intensity that of the marginally detected H$_2$ 2-1 S(2) line at 2.154~$\mu$m \citep[see Table~2 in ][]{Matsuura2007}. The H$_2$ excitation diagram of the K1 Helix's knot indicate a thermal origin for its H$_2$ emission and the dominant mechanism for the excitation of molecular hydrogen is UV-fluorescence pumping, rather than shocks \citep{Odell2000,Matsuura2007,Aleman_knots_2011,Andriantsaralaza2019}. We also included measurements made for a diffuse region  and a knot in Dumbbell by \citet{baldridge2017}. The comparable rotational and vibrational excitation temperatures inferred from a few lines indicates a thermal origin for the H$_2$ emission in Dumbbell nebula.

For NGC~7009, the $R$(H$_2$) ratios of LISs are determined $\sim$3.5 for the western LIS and around 9 for the eastern LISs. This difference in the $R$(H$_2$) ratio between the western and eastern LISs is attributed to the potential variation in the incident radiation field at their positions and/or in their column densities \citep{Sternberg1989,Burton1990,Aleman_knots_2011}.

The high $R$(H$_2$) ratios of the eastern LISs (between 8 and 10) indicate a thermal origin and the excitation mechanism can be either shock collisions or thermalized UV-pumping in a warm and high density ($>$10$^5$~cm$^{-3}$) H$_2$ gas \citep[e.g.][]{Sternberg1989,Burton1990}. As discussed above, for such values, it is not possible to disentangle the two mechanisms only from the $R$(H$_2$) ratio. The values of $R$(H$_2$) for the eastern LISs are comparable with those measured for two cometary knots in Helix and Dumbbell, and those in K~4-47 (Figure~\ref{plot}). In K~4-47, its H$_2$ emission is attributed to a thermalized shock-heated gas because of the high expansion velocities of the LISs \cite[$>$100\kms{}; ][]{Corradi2000} and high $R$(H$_2$) ratio \citep{Akras2017}. Although, thermal emission from pumped-H$_2$ molecules in warm and dense gas cannot be rule out. The western LIS exhibits $R$(H$_2$)$\sim$3, which is indicative of UV radiation determining the H$_2$ level population. This ratio is comparable to the values found for Hb~12.

A further comparison between the observed H$_2$ 1-0 S(1) and 2-1 S(1) emission-lines intensities and the theoretical predictions from photodissociation region (PDR) models \citep[see Table~5 in ][]{Burton1990} shows a very good agreement for high-density models. Models with densities of 10$^5$ and 10$^6$~cm$^{-3}$ provide surface brightness close to our observed values for the W1 and E1 LISs, respectively. The gas density difference indicates that the collisional deexcitation of pumped-H$_2$ is more efficient in E1 than in W1, resulting in more populated lower vibrational states and thus higher $R$(H$_2$) ratio.

Recall that long-slit low-resolution spectroscopy of NGC~7009 has shown that E1 and W1 LISs are also characterized by significantly different optical line ratios \citep[][]{Goncalves2003,Goncalves2006}. Based on 3D photoionization modelling, \citet[][]{Goncalves2006} demonstrated that two models with different optical depths between the LISs and the central star (or in other words altering the intensity of the ionising field of the central star at the distance of each LIS) are needed in order to reproduce their emission line fluxes. There is no clear reason why the optical depth or the photoionization rate between the eastern and western LISs should be different, given that both are located at approximately the same projected distance from the central star. Hence, the density of the gas in the LISs is more likely responsible for the divergences found in the line ratios.

For the pair of LISs (NE and SW) in NGC~6543, we determine $R$(Br$\gamma$) and lower limits for $R$(H$_2$). With the current data available, we cannot trace the excitation mechanism of the molecular gas in NGC~6543's LISs and spectroscopic follow-up observations are needed. UV-fluorescence emission \citep{Black1987,Sternberg1989,Burton1990} as well as thermal H$_2$ emission from shock-heated gas \citep{Hollenbach1989,Burton1992,Smith1995,Novikov2018} can explain the values obtained for the LISs in this nebula (Figure~\ref{plot}).

Besides the detection of H$_2$ emission in the LISs of NGC~6543, \citet[][]{Fang2018} also reported the detection of H$_2$ emission from several clumpy structures in the halo of the PN. Because of the strong optical \sulfurt\ and \nitrogen\ emission-lines detected in these structures, it was claimed that H$_2$ gas was shock-excited. However, strong \sulfurt\ and \nitrogen\ lines are not necessarily indicative of a shock-heated gas in PNe \cite{Akras2020}. 

\section{Conclusions}
We presented new very deep narrow-band H$_2$~1-0~S(1) and H$_2$~2-1~S(1) images of NGC~7009, and H$_2$~1-0~S(1) image of NGC~6543. H$_2$ emission was detected from the low-ionization structures in both PNe. The surface brightness of the H$_2$ 1-0 line was estimated between 0.46 and 2.9$\times$10$^{-4}$~erg~s$^{-1}$~cm$^{-2}$~sr$^{-1}$ for the LISs in NGC~7009 and between 0.29 and 0.48$\times$10$^{-4}$~erg~s$^{-1}$~cm$^{-2}$~sr$^{-1}$ for the LISs in NGC~6543, with S/N ratios between 10 and 42, and between 3 and 4, respectively. This findings increased the number of LISs with H$_2$ from four to six, further supporting the scenario that LISs are dense microstructures embedded in PNe, partially consisting of molecular gas.

Scrutinizing the $R$(H$_2$) and $R$(Br$\gamma$) diagnostic diagram, we concluded that the positive trend does not indicate a smooth transition from UV-fluorescent-dominated regions to shock-dominated regions. Based on the theoretical predictions from shock and UV-irradiated cloud models, both mechanisms are able to produce either low or high ratios. Therefore, both mechanisms are important in PNe and they contribute to the molecular hydrogen excitation. However, it is not feasible to trace the mechanism from this diagnostic diagram.

Regarding the NGC~7009, the high R(H$_2$) ratios ($\sim$8-10) derived from the eastern LISs were attributed either to a warm and dense gas with thermally populated states or to a thermalized shock-heated gas, whilst the low $R$(H$_2$) and $R$(Br$\gamma$) line ratios determined for the western LIS imply a non-thermal UV-fluorescence H$_2$ emission from a low-density gas.

On the other hand, the no detection of H$_2$ 2-1 emission from the LISs of NGC~6543 and the lower limit values obtained for $R$(H$_2$) ratio did not allow us to trace the origin of its emission. Radiative UV-fluorescence emission is more likely responsible for the excitation of H$_2$ gas, but shocks cannot be ruled out.

Overall, these new detections further confirmed the presence of molecular gas in high density, LISs in PNe. Although, a systematic H$_2$ imaging and spectroscopic survey of PNe with LISs is required in order to understand how the molecular hydrogen was formed or how it survived from the intense UV radiation and what is its dominant excitation mechanism.

\section*{Acknowledgements}
The authors would like to thank the anonymous referee for his/her thorough revision of the paper and his/her constructive comments and suggestions that improved the manuscript. SA acknowledges support from the Brazilian agencies CNPq (grant 454794/2015-0) and CAPES (PNPD fellowship). DRG is partially supported by CNPq grants 304184/2016-0 and 428330/2018-5. GR-L acknowledges support from CONACYT (Mexico). IA acknowledges the support of CAPES, Ministry of Education, Brazil, through a PNPD fellowship. The observations we report here were obtained at the Gemini Observatory (processed using the Gemini {\sc iraf} package and Gemini-{\sc python}), which is operated by the Association of Universities for Research in Astronomy, Inc., under a cooperative agreement with the NSF on behalf of the Gemini partnership: the National Science Foundation (United States), the National Research Council (Canada), CONICYT (Chile), the Australian Research Council (Australia), Minist\'{e}rio da Ci\^{e}ncia, Tecnologia e Inova\c{c}\~{a}o (Brazil), and Ministerio de Ciencia, Tecnolog\'{i}a e Innovaci\'{o}n Productiva (Argentina). This research is also based on observations made with the NASA/ESA {\it HST}, and obtained from the Hubble Legacy Archive, which is a collaboration between the Space Telescope Science Institute (STScI/NASA), the Space Telescope European Coordinating Facility (ST-ECF/ESA), and the Canadian Astronomy Data Centre (CADC/NRC/CSA).






\bibliographystyle{mnras}  
\bibliography{references}   


\bsp	
\label{lastpage}
\end{document}